\documentclass[12pt,a4paper]{article}
\usepackage[T1]{fontenc}
\usepackage[latin9]{inputenc}
\usepackage{float}
\usepackage{amsmath}
\usepackage{graphicx}
\usepackage{amssymb}

\usepackage{epsfig} 

\newcommand\plb[3]   {{\it Phys.\ Lett.\ }{\bf B #1} (#2) #3}
\newcommand\prd[3]   { {{\it Phys.\ Rev.\ }{\bf D #1} (#2) #3}} 
\newcommand\jhep[3]  {	{{\it J. High Energy Phys.\ }{\bf #1} (#2) #3}}
\newcommand\npb[3]   {{\it Nucl.\ Phys.\ }{\bf B #1} (#2) #3}
\newcommand\prl[3]   {{\it Phys.\ Rev.\ Lett.\ }{\bf #1} (#2) #3}
\newcommand\epjc[3]  {{\it Eur.\ Phys.\ J. }{\bf C #1} (#2) #3}
\newcommand\jphg[3]  {{\it J. Phys.\ }{\bf G #1} (#2) #3}

\begin{document}

\makeatother

\title{\textbf{\Large Electroweak Precision Constraints on Vector-like Fermions}}

\author{G. Cynolter and E. Lendvai}

\date{
Theoretical Physics Research Group of Hungarian Academy of
Sciences, E\"otv\"os University, Budapest, 1117 P\'azm\'any P\'eter s\'et\'any 1/A,
Hungary}

\maketitle

\begin{abstract}
We calculate the oblique electroweak corrections and confront with
the experiments in an extension of the Standard Model. The new fields
added are a vector-like weak doublet and a singlet fermion. After
electroweak symmetry breaking there is a mixing between the components
of the new fields, but no mixing allowed with the standard fermions.
Four electroweak parameters, $\hat{S}$, $\hat{T}$, $W$, $Y$ are
presented in the formalism of Barbieri et al., these are the generalization
of the Peskin-Takeuchi $S,\, T,\, U$'s. The vector-like extension
is slightly constrained, $\hat{T}$ requires the new neutral fermion
masses not to be very far from each other, allowing
higher mass difference for higher masses and smaller mixing. $\hat{S},\,W,\, Y$
gives practically no constraints on the masses.
This extension can give a positive contribution to $\hat{T} $, allowing a heavy Higgs boson in electroweak precision tests of the Standard Model. 
\end{abstract}


\section{Introduction}

Vector-like fermions appear in several extensions of the Standard
Model (SM). They are present in extra dimensional models with bulk
fermions e.g \cite{UED}, in little Higgs theories \cite{littleH},
in models of so called improved naturalness consistent with a heavy
Higgs scalar \cite{improved}, in simple fermionic models of dark
matter \cite{darkmatter,dmhiggs}, in some dynamical models of supersymmetry
breaking using gauge mediation, topcolor models \cite{top}, and were
also considered as the solution to the discrepancy between $R_{b}$
and LEP2 measurements in the mid 90's \cite{Rb}. Vector-like fermions
were essential ingredients in a recent proposal, in which a nontrivial
condensate of new vector-like fermions breaks the electroweak symmetry
and provides masses for the standard particles \cite{fcm}. The potential
LHC signals of vector-like quarks were discussed in \cite{jetmass}.

Any extension of the SM must face the tremendous success of the SM
in high energy experiments, it must have evaded direct detection and
fulfill the electroweak precision tests. If the scale of new physics
is sufficiently high and the corrections are assumed to be universal
then the new physics only affects the finite combinations of the gauge
boson self-energies. These parameters (traditionally S,T,U \cite{stu})
are constrained by experiments \cite{pdg}. Barbieri et al. reconsidered
the problem \cite{lep2} and showed that there are indeed four relevant
parameters $\hat{S},\,\hat{T},\, W,\, Y$, where $\hat{S}$ and $\hat{T}$
are related to the old parameters $S=4s_{W}^{2}\hat{S}/\alpha$, $T=\hat{T}/\alpha$.
$W$ and $Y$ are two new parameters, $U$ ($\hat{U}$) is suppressed
by the scale of new physics compared to $T$ ($\hat{T}$). There are
also other and more extended parameterizations are known \cite{stu6}.

In this letter we calculate the gauge boson vacuum polarization functions
and precision electroweak observables for vector-like extensions of
the SM, especially taking into account the mixing in the recently
proposed fermion condensate model \cite{fcm}. There are earlier results
for extra vector-like quarks \cite{silva,dmhiggs} and detailed calculations
for the $\rho$ parameter in the littlest Higgs model e.g. \cite{dawson}.
The new results in this paper are that we use different representations
for the new fermions, give general formulae applicable to LEP2 measurements
using the four parameters of \cite{lep2} and constrain the fermion
condensate model \cite{fcm}.

\section{Extension of the Standard Model with \protect \\
vector-like fermions}

We consider a simple extension of the SM based on non-chiral fermions.
The new colorless fermions are an extra neutral weak $SU(2)$ singlet
$\Psi_{S}$ ($T=Y=0$) and a doublet $\Psi_{D}=\left(\begin{array}{c}
\Psi_{D}^{+}\\ \Psi_{D}^{0}\end{array}\right)$ 
with hypercharge 1. It is assumed that the new fermions  are odd under
a new $Z_{2}$ symmetry, while all the standard particles are even.
This symmetry forbids mixings with standard fermions and the lightest
new fermion is stable providing an ideal weakly interacting dark matter
candidate. The purely fermionic part of the new Lagrangian is\begin{equation}
L_{\Psi}=i\overline{\Psi}_{D}D_{\mu}\gamma^{\mu}\Psi_{D}+i\overline{\Psi}_{S}\partial_{\mu}\gamma^{\mu}\Psi_{S}-m_{1}\overline{\Psi}_{D}\Psi_{D}-m_{2}\overline{\Psi}_{S}\Psi_{S},\label{eq:kinetic}\end{equation}
with Dirac masses $m_{1},m_{2}$. $\Psi_{S}$ may have further interactions
irrelevant for our analysis. $D_{\mu}$ is the covariant derivative
\begin{equation}
D_{\mu}=\partial_{\mu}-i\frac{g}{2}\underline{\tau}\,\underline{W}_{\mu}-i\frac{g'}{2}B_{\mu},\label{eq:covariantd}\end{equation}
where $\underline{W}_{\mu,}B_{\mu}$ and $g,\; g'$ are the usual
weak gauge boson fields and couplings, respectively. In a renormalizable
theory including the standard Higgs doublet ($H$) additional Yukawa
terms appear resulting a mixing between the new neutral fermions.\begin{equation}
L_{Yukawa}=\lambda_{m}\overline{\Psi}_{D}\Psi_{S}H+\lambda_{m}^{*}H^{\dagger}\overline{\Psi}_{S}\Psi_{D}\label{eq:Higgs yukawa}\end{equation}
In a version of the Standard Model \cite{fcm}, the Higgs boson is
a composite state of the new fermions ($H=\overline{\Psi}_{S}\Psi_{D}$)
and these Yukawa terms (and additional contribution to $\Psi_{D},\ \Psi_{S}$
mass ) generated by condensation from effective 4-fermion interactions.

When the Higgs (or the composite operator $\overline{\Psi}_{S}\Psi_{D}$
in \cite{fcm}) develops a vacuum expectation value, $\left\langle H\right\rangle _{0}=\left(\begin{array}{c}
0\\
v\end{array}\right)$, with real $v$, non-diagonal mass terms are generated with $m_{3}=(\lambda_{m}+\lambda_{m}^{*})v/2$
\begin{equation}
L_{\hbox{mass}}=-m_{1}\overline{\Psi}_{D}\Psi_{D}-m_{2}\overline{\Psi}_{S}\Psi_{S}-m_{3}\left(\overline{\Psi^{0}}_{D}\Psi_{S}+\overline{\Psi}_{S}\Psi_{D}^{0}\right).\label{eq:mass}
\end{equation}
In \cite{fcm} $m_{1}(m_{2})$ get contributions from the condensates.
The mass matrix of the new fermions must be diagonalized via unitary
transformation to get physical mass eigenstates 

\begin{eqnarray}
\Psi_{1} & = & \phantom{-}c\,\Psi_{D}^{0}+s\,\Psi_{S},\nonumber \\
\Psi_{2} & = & -s\,\Psi_{D}^{0}+c\,\Psi_{S},\label{eq:fermion mix}\end{eqnarray}
where $c=\cos\phi$, $s=\sin\phi$, $\phi$ is the mixing angle defined
by\begin{equation}
2m_{3}=(m_{1}-m_{2})\tan2\phi.\label{eq:def phi}\end{equation}
The masses of the new neutral physical fermions $\Psi_{1},\:\Psi_{2}$
are $M_{1,2}=\frac{1}{2} \left ( m_{1}+m_{2}\pm\frac{m_{1}-m_{2}}{\cos2\phi} \right ).$
The useful inverse relations are 

\begin{align}
m_{1}= & c^{2}M_{1}+s^{2}M_{2},\nonumber \\
m_{2}= & s^{2}M_{1}+c^{2}M_{2}.\label{eq:mphys}\end{align}

In the physical spectrum there is also a charged fermion $\Psi_{D}^{+}$,
with mass $M_{+}=m_{1}$ (given by (\ref{eq:mphys})). In the case
of an elementary scalar field $\lambda_{m}$ is a free parameter.
The mixing angle and the physical masses are basicly not constrained
from the theory. In \cite{fcm} gap equations determine the masses
and the mixing angle. Applying further unitarity constraints one finds
that one of the neutral masses is very close the charged mass and
the mixing is rather weak \cite{fcmgap}.

The collider phenomenology and radiative corrections in the model
are coming from the doublet kinetic term in (\ref{eq:kinetic}) taking
into account the mixing (\ref{eq:fermion mix})

\begin{eqnarray}
L^{I} & = & \phantom{+}\overline{\Psi_{D}^{+}}\gamma^{\mu}\Psi_{D}^{+}\left(\frac{g'}{2}B_{\mu}+\frac{g}{2}W_{3\mu}\right)+\nonumber \\
 &  &
 +\left(c^{2}\overline{\Psi}_{1}\gamma^{\mu}\Psi_{1}+s^{2}\overline{\Psi}_{2}\gamma^{\mu}\Psi_{2}-sc 
\left(\overline{\Psi}_{1}\gamma^{\mu}\Psi_{2}+\overline{\Psi}_{2}\gamma^{\mu}\Psi_{1}\right)\right)\left(\frac{g'}{2}B_{\mu}-\frac{g}{2}W_{3\mu}\right)+\nonumber \\
 &  &
 +\left[\frac{g}{\sqrt{2}}W_{\mu}^{+}\left(c\overline{\Psi_{D}^{+}}\gamma^{\mu}\Psi_{1}-s
\overline{\Psi_{D}^{+}}\gamma^{\mu}\Psi_{2}\right)+h.c.\right].\label{eq:Llmix}
\end{eqnarray}
We calculate the contribution to the electroweak precision observables
from this renormalizable interaction.

\section{Electroweak precision parameters}

Barbieri et al.  showed \cite{lep2} that if the scale of new physics is sufficiently higher than the LEP2 scale
and the new physics affects only the vector boson self energies then
the most general parameterization
of new physics effects uses 4 parameters $\hat{S},\,\hat{T},\, W,\, Y$.
These parameters are the generalizations of the Peskin-Takeuchi S,T,U
parameters and defined from the transverse gauge boson vacuum polarization
amplitudes \begin{equation}
\Pi_{ab}^{\mu\nu}(q^{2})=g^{\mu\nu}\Pi_{ab}(q^{2})+p^{\mu}p^{\nu}\hbox{terms}\label{eq:pidef}\end{equation}
expanded up to the quadratic order 
$\left(ab=\{W^{+}W^{-},\ W_{3}W_{3},\ BB,\ W_{3}B\}\right)
$\[
\Pi_{ab}(q^{2})\simeq\Pi_{ab}(0)+q^{2}\Pi'_{ab}(0)+\frac{\left(q^{2}\right)^{2}}{2}\Pi''_{ab}(0)+...\] \,.
The relevant parameters are defined by\begin{eqnarray}
(g'/g)\hat{S} & = & \Pi'_{W_{3}B}(0),\label{eq:shat}\\
M_{W}^{2}\hat{T} & = & \Pi_{W_{3}W_{3}}(0)-\Pi_{W^{+}W^{-}}(0),\label{eq:that}\\
2M_{W}^{-2}Y & = & \Pi''_{BB}(0),\label{eq:yhat}\\
2M_{W}^{-2}W & = & \Pi''_{W_{3}W_{3}}(0),\label{eq:what}\end{eqnarray}
here we use canonically normalized fields and $\Pi$ functions. The
form factor $\hat{T}$ has custodial and $SU_{L}(2)$ breaking quantum
numbers, while $\hat{S}$ respects custodial symmetry and breaks $SU_{L}(2)$.
$Y$ and $W$ are symmetric under both symmetries and they are important
at the LEP2 energies. The result of the combined fit (excluding NuTeV)
is shown in Table 1. from \cite{lep2}.

\begin{table}[H]
\begin{center}
\begin{tabular}{|c|c|c|c|c|}
\hline 
 & $10^{3}\hat{S}$ & $10^{3}\hat{T}$ & $10^{3}W$ & $10^{3}Y$\tabularnewline
\hline
light Higgs & 0.0 $\pm$1.3 & 0.1 $\pm0.9$ & 0.1$\pm$1.2 & -0.4$\pm$0.8\tabularnewline
\hline 
heavy Higgs & -0.9$\pm$1.3 & 2.0$\pm$1.0 & 0.0$\pm$1.2 & -0.2 $\pm$0.8\tabularnewline
\hline
\end{tabular}
\par\end{center}

\caption{Global fit of the electroweak precision parameters for a light ($M_{H}=115$
GeV) and a heavy ($M_{H}=800$ GeV) Higgs.}

\end{table}

The calculation of the parameters is based on the general gauge boson
vacuum polarization diagram with two non-degenerate fermions with
masses $m_{a}$ and $m_{b}$. We use dimensional regularization and
give the result for general $q^{2}$.
The coupling constants are defined in the usual manner $L^{I}\sim V_{\mu}\bar{\Psi}\left(g_{V}\gamma^{\mu}+g_{A}\gamma_{5}\gamma^{\mu}\right)\Psi$.

\begin{equation}
\Pi(q^{2})=\frac{1}{4\pi^{2}}\left(\left(g_{V}^{2}+g_{A}^{2}\right)\tilde{\Pi}_{V\!+\! A}+\left(g_{V}^{2}-g_{A}^{2}\right) \tilde{\Pi}_{V\!-\! A}\right)\label{eq:pivv}
\end{equation}
\begin{eqnarray}
\tilde{\Pi}_{V\!+\! A} & = & -\frac{1}{2}\left(m_{a}^{2}+m_{b}^{2}-\frac{2}{3}q^{2}\right)
\left( \hbox{Div} +\ln\left(\frac{\mu^{2}}{m_{a}m_{b}}\right) \right )
-\frac{\left(m_{a}^{2}-m_{b}^{2}\right)^{2}}{6q^{2}}-\frac{1}{3}\left(m_{a}^{2}+m_{b}^{2}\right)+\label{eq:piv+a}\\
 &  &
 +\frac{5}{9}q^{2}-\frac{\left(m_{a}^{2}-m_{b}^{2}\right)^{3}}{12q^{4}}\ln\left(\frac{m_{b}^{2}}{m_{a}^{2}}\right)
+\frac{1}{3} \left(\frac{\left(m_{a}^{2}-m_{b}^{2}\right)^{2}}{q^{2}}+m_{a}^{2}+m_{b}^{2}-2q^{2}\right)f
\left(m_{a}^{2},m_{b}^{2},q^{2}\right) \nonumber 
\end{eqnarray}
and 
\begin{equation}
\tilde{\Pi}_{V\!-\!A}=m_{a}m_{b}\left(\hbox{Div}+\ln\left(\frac{\mu^{2}}{m_{a}m_{b}}\right)  +2+
\frac{\left(m_{a}^{2}-m_{b}^{2}\right)}{2q^{4}}\ln\left(\frac{m_{b}^{2}}{m_{a}^{2}}\right)-2f\left(m_{a}^{2},m_{b}^{2},q^{2}\right)\right).\label{eq:piv-a}
\end{equation}
The function $f\left(m_{a}^{2},m_{b}^{2},q^{2}\right)$ is given by
\begin{equation}
f\left(m_{a}^{2},m_{b}^{2},q^{2}\right)=\left\{ \begin{array}{cc}
\phantom{-}\sqrt{\Delta}\hbox{arctanh}^{-1}\left(\frac{\sqrt{\Delta}q^{2}}{q^{2}-\left(m_{a}+m_{b}\right)^{2}}\right) & q<\left|m_{a}-m_{b}\right|\\
-\sqrt{-\Delta}\arctan\left(\frac{\sqrt{-\Delta}q^{2}}{q^{2}-\left(m_{a}+m_{b}\right)^{2}}\right) & \left|m_{a}-m_{b}\right|<q\ and\ q<m_{a}+m_{b}\\
\phantom{-}\sqrt{\Delta}\hbox{arccot}^{-1}\left(\frac{\sqrt{\Delta}q^{2}}{q^{2}-\left(m_{a}+m_{b}\right)^{2}}\right) & m_{a}+m_{b}<q\end{array},\right.\label{eq:f}\end{equation}
where we defined
\begin{equation}
\Delta=1-2\frac{m_{a}^{2}+m_{b}^{2}}{q^{2}}+\frac{\left(m_{a}^{2}-m_{b}^{2}\right)^{2}}{q^{4}},\label{eq:delta}
\end{equation}
and $\hbox{Div}=1/\epsilon+\ln4\pi-\gamma_{\epsilon}$
contains the usual divergent term in dimensional regularization.

The electroweak parameters depend on the values and derivatives of
the $\Pi$ functions at $q^{2}=0$, the limits are given below.
\begin{eqnarray}
\tilde{\Pi}_{V\!+\! A}(0) & = & -\frac{1}{2}\left(m_{a}^{2}+m_{b}^{2}\right)
\left (\hbox{Div} +\ln\left(\frac{\mu^{2}}{m_{a}m_{b}}\right) \right )-  \\
& & -\frac{1}{4} (m_{a}^{2}+m_{b}^{2} )-\frac{\left(m_{a}^{4}+m_{b}^{4}\right)}{4\left(m_{a}^{2}-m_{b}^{2}\right)}
\ln\left(\frac{m_{b}^{2}}{m_{a}^{2}}\right)\label{eq:piv+a0}  \nonumber \\
\tilde{\Pi}_{V\!-\! A}(0) & = & m_{a}m_{b}\left(\hbox{Div}+\ln\left(\frac{\mu^{2}}{m_{a}m_{b}}\right) +1+\frac{\left(m_{a}^{2}+m_{b}^{2}\right)}{2\left(m_{a}^{2}-m_{b}^{2}\right)}\ln\left(\frac{m_{b}^{2}}{m_{a}^{2}}\right)\right)\label{eq:piv-a0}\end{eqnarray}
The first and second derivatives are\begin{eqnarray}
\tilde{\Pi}'_{V\!+\! A}(0) & \!\!=\!\!\! & 
\left ( \frac{1}{3}\hbox{Div} +\ln\left(\frac{\mu^{2}}{m_{a}m_{b}}\right) \right ) \!+\!\frac{m_{a}^{4}-8m_{a}^{2}m_{b}^{2}+m_{b}^{4}}{9\left(m_{a}^{2}-m_{b}^{2}\right)^{2}}+ \label{eq:piv+av0} \\ & & +\frac{\left(m_{a}^{2}+m_{b}^{2}\right)\left(m_{a}^{4}-4m_{a}^{2}m_{b}^{2}+m_{b}^{4}\right)}{6\left(m_{a}^{2}-m_{b}^{2}\right)^{3}}\ln\left(\frac{m_{b}^{2}}{m_{a}^{2}}\right)\nonumber \\
\tilde{\Pi}'_{V\!-\! A}(0) & = & m_{a}m_{b}\left(\frac{\left(m_{a}^{2}+m_{b}^{2}\right)}{2\left(m_{a}^{2}-m_{b}^{2}\right)}+\frac{m_{a}^{2}m_{b}^{2}}{\left(m_{a}^{2}-m_{b}^{2}\right)^{3}}\ln\left(\frac{m_{b}^{2}}{m_{a}^{2}}\right)\right),\label{eq:piv-av0}\end{eqnarray}
and \begin{eqnarray}
\tilde{\Pi}''_{V\!+\! A}(0) & = & \frac{\left(m_{a}^{2}+m_{b}^{2}\right)\left(m_{a}^{4}-8m_{a}^{2}m_{b}^{2}+m_{b}^{4}\right)}{4\left(m_{a}^{2}-m_{b}^{2}\right)^{4}}-\frac{3m_{a}^{4}m_{b}^{4}}{\left(m_{a}^{2}-m_{b}^{2}\right)^{5}}\ln\left(\frac{m_{b}^{2}}{m_{a}^{2}}\right),\label{eq:piv+avv0}\\
\tilde{\Pi}''_{V\!-\! A}(0) & = & m_{a}m_{b}\left(\frac{\left(m_{a}^{4}+10m_{a}^{2}m_{b}^{2}+m_{b}^{4}\right)}{3\left(m_{a}^{2}-m_{b}^{2}\right)^{4}}+\frac{2\left(m_{a}^{2}+m_{b}^{2}\right)m_{a}^{2}m_{b}^{2}}{2\left(m_{a}^{2}-m_{b}^{2}\right)^{5}}\ln\left(\frac{m_{b}^{2}}{m_{a}^{2}}\! \right)\!  \right).\label{eq:piv-avv0}
\end{eqnarray}

The values of the vacuum polarizations for identical masses $(m_{b}=m_{a})$
are the smooth limits of the previous formulae and agree with direct
calculation.\[
\begin{array}{ll}
\tilde{\Pi}{}_{V\!+\! A}(0)=-m_{a}^{2} \hbox{Div}-m_{a}^{2} \ln \left(\frac{\mu^{2}}{m_{a}^2}\right), 
 & \tilde{\Pi}{}_{V\!-\! A}(0)=m_{a}^{2}\hbox{Div} +m_{a}^{2} \ln \left(\frac{\mu^{2}}{m_{a}^2}\right),\\
\tilde{\Pi}'_{V\!+\! A}(0)=\frac{1}{3}\hbox{Div}+\frac{1}{3}m_{a}^{2} \ln\left(\frac{\mu^{2}}{m_{a}^2}\right) -\frac{1}{6},\quad  & \tilde{\Pi}'_{V\!-\! A}(0)=\frac{1}{6},\\
\tilde{\Pi}''_{V\!+\! A}(0)=\frac{1}{10m_{a}^{2}}, & \tilde{\Pi}''_{V\!-\! A}(0)=\frac{1}{30m_{a}^{2}}.\end{array}\]

The new vector-like fermions contribute to the complete
vacuum polarization as the sum of (\ref{eq:piv+a}) and (\ref{eq:piv-a}).
We define
\begin{equation}
\tilde{\Pi}_V(m_{a},m_{b},q^{2})=\tilde{\Pi}_{V\!+\! A}(m_{a},m_{b},q^{2})+
 \tilde{\Pi}_{V\!-\! A}(m_{a},m_{b},q^{2}).\label{eq:def pitilde}
\end{equation}
In what follows the index $V$ is omitted we use $\tilde{\Pi}=\tilde{\Pi}_V $.

The $\hat{S}$ parameter (\ref{eq:shat}) is then given by 
\begin{equation}
\hat{S}=\frac{g^{2}}{16\pi^{2}}\left(+\tilde{\Pi}'(M_{+},M_{+},0) -c^{4}\tilde{\Pi}'(M_{1},M_{1},0) -s^{4}\tilde{\Pi}'(M_{2},M_{2},0) -2s^{2}c^{2}\tilde{\Pi}'(M_{2},M_{1},0)\right).\label{eq:shat par}
\end{equation}
The first three terms cancel the divergent contribution of the last one.

The $\hat{T}$ parameter (\ref{eq:that}) related to $\Delta\rho$ is also finite.
\begin{eqnarray}
\hat{T} & = & \frac{g^{2}}{M_{W}^{2}16\pi^{2}}\left[+\tilde{\Pi}(M_{+},M_{+},0) +c^{4}\tilde{\Pi}(M_{1},M_{1},0) +s^{4}\tilde{\Pi}(M_{2},M_{2},0)+\right.\nonumber \\
 &  &  \left. 
+2s^{2}c^{2}\tilde{\Pi}(M_{2},M_{1},0) -2c^{2}\tilde{\Pi}(M_{+},M_{1},0) -2s^{2}\tilde{\Pi}(M_{+},M_{2},0)\right]
.\label{eq:that par}
\end{eqnarray}
The $Y$ and the $W$ parameters differ only in the coupling constants\begin{eqnarray}
Y & = & M_{W}^{2}\frac{g'^{2}}{32\pi^{2}}\cdot\left[\tilde{\Pi''}(M_{+},M_{+},0)+c^{4}\tilde{\Pi}''(M_{1},M_{1},0)+\right.\nonumber \\
 &  &
 \left.\phantom{M_{W}^{2}\frac{g'^{2}}{2\pi^{2}}}+s^{4}\tilde{\Pi}''(M_{2},M_{2},0)
+2s^{2}c^{2} \tilde{\Pi}''(M_{2},M_{1},0)\right],\label{eq:Y}\\
W & = & M_{W}^{2}\frac{g^{2}}{32\pi^{2}}\left[\tilde{\Pi''}(M_{+},M_{+},0)+c^{4}\tilde{\Pi}''(M_{1},M_{1},0)+\right.\nonumber \\
 &  & 
\left.\phantom{M_{W}^{2}\frac{g{}^{2}}{2\pi^{2}}}+s^{4}\tilde{\Pi}''(M_{2},M_{2},0)+2s^{2}c^{2}\tilde{\Pi}''(M_{2},M_{1},0)\right].\label{eq:W}
\end{eqnarray}
The first three terms in the parentheses give $W=\frac{g^2}{240\pi^2}M_W^2 \cdot \left(1/M_{+}^{2}+c^{4}/M_{1}^{2}+s^{4}/M_{2}^{2}\right)$ in agreement with \cite{Mar2005} taking into account that they considered Majorana fermions. The last term is the same order of magnitude in $M_W/M_{\{1,2,+ \}} $. Here $W$ and $Y$ are always non-negative fulfilling the positivity constraints proven in \cite{minimalset}.

\section{Numerical results}

\begin{figure}[t]
\begin{center}
\includegraphics[scale=0.50]{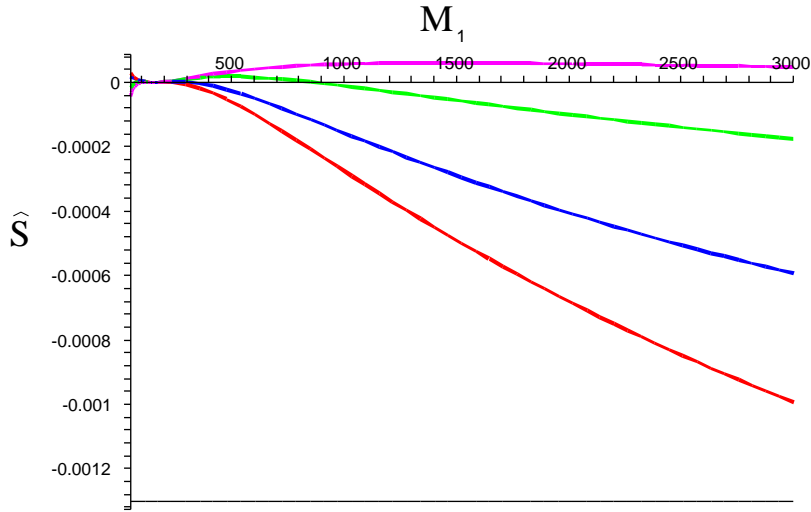}
\end{center}
\begin{center}
\caption{The $\hat{S}$ parameter vs. $M_{1}$ for $M_{2}=150$ 
GeV for $c^{2}=0.2,\,0.4,\,0.6,\,0.8$ respectively from bottom upwards
(from red to magenta), the horizontal line is the 1$\sigma$ experimental
lower bound.}

\end{center}
\end{figure}
There are 3 free parameter in the model to confront with the experiments:
the two neutral masses ($M_{1,2}$) and the mixing angle $ \phi $, $ s^2=\sin^2 \phi $, $c^{2}=\cos^{2}\phi$.
The mass of the charged fermion is given by $M_{+}=c^{2}M_{1}+s^{2}M_{2}$
(\ref{eq:mphys}). The new particles are expected to be heavier than
approximately $100$ GeV from LEP1 and LEP2 as they have ordinary couplings
with the gauge bosons. For relatively light  new particles (with masses 100-150 GeV)
the oblique parameters give rough estimate of the radiative corrections \cite{Mar2005}.
Replacing $M_{1}\leftrightarrow M_{2}$ and
$c^{2}\leftrightarrow s^{2}=1-c^{2}$ gives the same oblique parameters.
If there is no real mixing $c^{2}=0$ or $1$ or if $M_{1}=M_{2}=M_{+}$
then there is one degenerate vector-like fermion doublet and a decoupled
singlet, $\hat{S}$ and $\hat{T}$ vanish explicitely. In this case
the new sector does not violate $SU_{L}(2)$ and there is an exact
custodial symmetry. Increasing the mass difference in the remnants
of the original doublet by increasing the $\left|M_{1}-M_{2}\right|$
mass difference and/or moving away from the non-mixing case $c^{2}=0,$ or $1$
results in increasing $\hat{S}$ and $\hat{T}$. For small violation
of the symmetries $\hat{S}$ and $\hat{T}$ are expected to be small.

The $\hat{S},\, W,\, Y$ parameters are small for masses in the range
from 100 GeV up to few TeV, the only exception is $\hat{T}\,(T)$,
which is sensitive to the mass differences. These features were predicted
using simple assumptions in \cite{fcmgap}. We discuss in details
the case of a light Higgs boson (Table 1).

Generally the $\hat{S}\,(S)$ parameter depends only on the masses
of the new particles and the mixing angle, it contains no further
dimensional parameter. For reasonable masses (below few TeV) it is
always in agreement with the 1$\sigma$ experimental bounds for $M_{1,2}\geq100$
GeV. See Fig. 1. 
 For higher masses $\hat{\left|S\right|}\,(\left|S\right|)$
is even smaller.  $S$  can have both signs.
\begin{figure}
\begin{center}
\includegraphics[scale=0.50]{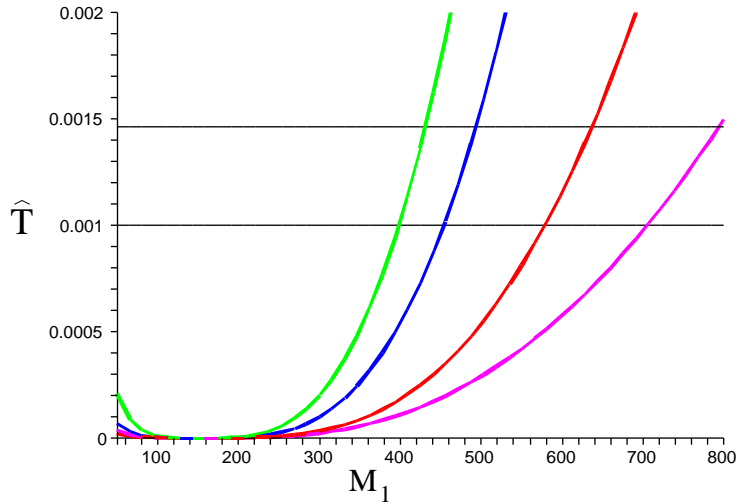}
\end{center}
\noindent \caption{The $\hat{T}$ parameter vs. $M_{1}$ for $M_{2}=150,$ GeV for
$c^{2}=0.9,\,0.1,\,0.2,\,0.55$ from bottom upwards, the horizontal
lines are the 1$\sigma$ and 1.6 $\sigma$ experimental upper bounds.}
\end{figure}
The $\hat{T}$ parameter (\ref{eq:that par})  is more sensitive
to the value of $M_{1,2}$. The mass difference of the new
fermions must not exceed  a critical value, $|M_1-M_2| \leq 250 \; (400) $  for the mass 
of the lighter fermion  150 (500) GeV.
The constraints are the strongest for $c^{2}\simeq 0.56$,
below and above this mixing the absolute value of $\hat{T}$ decreases.
For small mixing ($c^{2}$ close to 0 or 1) there are very weak or
simply no constraints. Fig. 2. shows as an example $\hat{T}$ for $M_{2}=150$
GeV as a function of $M_{1}$ for various mixings. $c^{2}=0\,(1)$
gives a horizontal line, $\hat{T}=0$.  
$\hat{T}\,(T)$  is always positive allowing a heavy Higgs particle.

The $W$ parameter is sensitive to the ratio $M_{W}^{2}/M_{i}^{2}$,
$i=1,2,+$. It is largest for relatively small masses approximately (150 GeV), 
but $W$ is still well within the  $ 1\sigma$ experimental limits.
For higher masses $W$ is even smaller. See Fig.3. 
 The $Y$ parameter is the same function of the masses and mixing angles as $W$. 
The smaller gauge coupling multiplier provides weaker constraints.

If the Higgs is heavy, $M_{H}=800$ GeV (see Table 1.) the central
value of $\hat{S}$ decreases compared to the light Higgs
case. $\hat{S}$  and $ W $ gives no constraints. 
At the same time the negative contribution of the light Higgs can be compensated by 
the new fermions with considerable mass difference for example (150,400) GeV or (500, 900) GeV. 
Non-degenerate vector-like fermions allow a space for heavy Higgs in the precision tests of the Standard Model.

\begin{figure}[t]
\begin{center}
\includegraphics[scale=0.50]{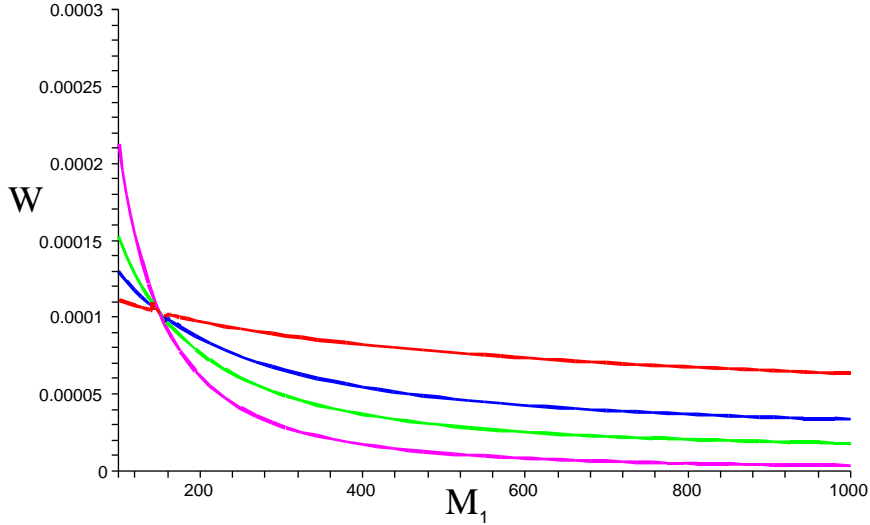}

\end{center}

\caption{The W parameter vs. $M_{1}$ for $M_{2}=150$ GeV for $c^{2}=0.1,\,0.3,\,0.5,\,0.9$
respectively from top  downwards at high $M_{1}$, 
 the 1$\sigma$ experimental  bound is at 0.0013 is outside the figure.}
\end{figure}
In the fermion condensate model \cite{fcm} the Higgs boson in \eqref{eq:Higgs yukawa}
is a composite state of the new fermions. Gap equations were derived
and solved for the parameters of the model. Applying further perturbative
unitarity arguments constrains the model seriously, the charged particle
mass must be relatively close to one of the neutral ones, e.g. $c^{2}$
must be close to $0$ or $1$ \cite{fcmgap}. The solutions of the
gap equations fulfill easily the experimental constraints on $\hat{S}$
and $\hat{T}$ due to the small mixing and the $ W$ parameter is also safe.
The solutions in \cite{fcmgap} result that the
oblique corrections do not constrain the fermion condensate model
even if the neutral masses $(M_{1,2})$ are non-degenerate. 
The calculation presented in this paper  shows that the fermion condensate model 
is less constrained than assumed by the naive estimates in \cite{fcmgap}.
The formulae derived here can be applied not just to \cite{fcm}, but
to various models generating the Lagrangian \eqref{eq:Higgs yukawa}.

\section{Conclusions}

We have calculated the oblique corrections in an extension of the
Standard Model based on vector-like weak singlet and doublet fermions.
Due to non-diagonal mass terms (\ref{eq:mass}) symmetry breaking
mixing occurs between the singlet and the neutral component of the
doublet. The oblique corrections were presented in the formalism of
Barbieri et al. \cite{lep2}. There are four relevant parameters $\hat{S},\,\hat{T},\, W,\, Y$,
and they are indeed in the same order of magnitude in the allowed
mass range, as expected. $Y$ is the same function of the masses and
mixing angle as $W$ with smaller coupling constant, but with weaker
constraints therefore we kept $\hat{S},\,\hat{T},\, W$. The corrections
depend on the new fermion masses ($M_{1,2}$) and the mixing angle.
The $\hat{S},\, W$ parameters are always in agreement with experiment for masses below few TeV.
The $\hat{T}\,(T)$ parameter
measures the custodial symmetry breaking, the custodial symmetry is
exact in the new sector if there is no physical mixing: $c^{2}=0,\,1$
or $M_{1}=M_{2}$. Depending on the mixing angle it allows in the
most stringent case for $c^{2}\simeq 0.56$ a maximal mass difference
$\left|M_{1}-M_{2}\right|\lesssim250$ GeV at $ 1 \sigma$  for relatively
small ligter neutral mass ($\sim150$ GeV), higher mass difference is allowed for
higher $M_{1,2}$ masses or differen mixing. This extension/modification
nicely accommodates a heavy Higgs in the Standard Model. The lightest new fermion
is stable and a good dark matter candidate. The model can be tested
at LHC in the Drell-Yan process \cite{fcm} or via jetmass analysis
\cite{jetmass}. Nearing the completion of our work we received a
preprint which deals with similar topic, but with different fermion
representation, approach and mixing allowed with the standard fermions
\cite{leptonspain}.

\subsubsection*{Acknowledgment}

The authors thank George P\'ocsik for valuable discussion.

\bibliography{EFHPP}
\bibliographystyle{JHEP}

\providecommand{\href}[2]{#2}\begingroup\raggedright\endgroup

\end{document}